\documentstyle[prb,eqsecnum,preprint,aps]{revtex}
                                                                     
\begin{document}

\title{Nonlocal conductivity in the vortex liquid regime of a
two-dimensional superconductor}

\author{Rachel Wortis$^{1,2}$ and David A. Huse$^{2,3}$}

\address{
$^1$Department of Physics, 
 University of Illinois at Urbana-Champaign, \\
1110 West Green Street, Urbana, Illinois 61801 }

\address{
$^2$Bell Laboratories, Murray Hill, NJ 07974 }

\address{
$^3$  Address after August 1996: Physics Dept., Princeton Univ.,
Princeton, NJ 08544 }

\date{\today}

\maketitle

\begin{abstract}

We have simulated the time-dependent Ginzburg-Landau equation with
thermal fluctuations, to study the nonlocal dc conductivity of a 
superconducting film.  
Having examined points in the phase diagram at a wide range of
temperatures and fields below the mean-field upper critical field,
we find a portion of the vortex-liquid regime
in which the nonlocal ohmic conductivity in real space is 
negative over a distance several times the spacing between vortices.
The effect is suppressed when driven beyond linear response.
Earlier work had predicted the existence of such a regime,
due to the high viscosity of a strongly-correlated vortex liquid.
This behavior is clearly distinguishable from the monotonic 
spatial fall-off of the conductivity in the higher temperature
or field regimes approaching the normal state.  The possibilities 
for experimental study of the nonlocal transport properties
are discussed.

\end{abstract}

\section{Introduction}

In this paper we present the results of a computer simulation 
designed to study the nonlocal dc conductivity of a 
two-dimensional (thin film) type-II superconductor.  
The meaning of nonlocal in this context can be seen in the 
standard expression connecting the local current density in a 
material, ${\bf J}$, to the local electric field, ${\bf E}$, 
in the linear (Ohmic) regime:
\begin{equation}
J_{\mu}({\bf r})=\int \sigma_{\mu \nu}({\bf r},{\bf r'})
E_{\nu}({\bf r'}) d{\bf r'} .
\end{equation}
When the conductivity, $\sigma ({\bf r},{\bf r '})$, is nonzero 
for ${\bf r} \neq {\bf r '}$, then it is nonlocal.  In a 
translationally invariant system, the nonlocal conductivity can 
only be a function of the 
difference $({\bf r} -{\bf r '})$.  The Fourier transform of the 
conductivity equation is then 
$J_{\mu}({\bf k})=\sigma_{\mu \nu}({\bf k}) E_{\nu}({\bf k})$. 
The nonlocal conductivity we are discussing here is a different
phenomenon from Pippard's nonlocal relation between the supercurrent
and the vector potential below $T_c$ in a type-I 
superconductor.\cite{tink}
In particular, we are interested in the nonlocal conductivity
in the resistive vortex-liquid regime of a type-II superconductor
in a magnetic field.

All materials exhibit nonlocal transport properties on some length
scale.  Normal metals behave nonlocally on length scales less than or
of the order of the inelastic mean free path.  In superconductors,
however, the scale of the nonlocality can be much larger.  Israeloff,
{\it et al.}\cite{isr} have measured effects arising from the 
nonlocal resistivity due to superconducting fluctuations just above 
$T_c$ in one dimensional rings of type-I material.  
The observed behavior, predicted by Glazman, {\it et
al.},\cite{glazman} arises from the correlations of the 
superconducting order parameter. 

In the mixed state of a type-II superconductor, the presence of 
vortices provides another mechanism for nonlocal resistivity.
When a current ${\bf J}({\bf r '})$ exerts Lorentz and Magnus 
forces on a vortex segment at ${\bf r '}$ and causes it to move, 
this vortex motion can in turn cause vortex segments at ${\bf r}$ 
to move, through vortex interactions, connections or entanglement.  
The motion of the vortices at ${\bf r}$ produces phase-slip and 
electric fields at ${\bf r}$, as described by the Josephson
relation, completing a nonlocal relation between current and 
electric field.  Such nonlocal resistivity over length scales 
of tens of microns
has been observed in bulk crystals of the high-temperature
superconductor YBCO by Safar, {\it et al}.\cite{safar} and
discussed phenomenologically by Huse and Majumdar.\cite{h&m}

In a recent paper with Mou and Dorsey,\cite{us} we 
examined the nonlocal dc transport properties
throughout the phase diagram of a type-II superconductor, using
analytic calculations where possible and proposing phenomenological
arguments elsewhere.  In particular, we predicted that the 
wavevector-dependent dc electrical conductivity, $\sigma(k)$, 
of a type-II superconductor would have a nonmonotonic dependence on
$k$ in a certain region of the phase diagram:
For those values of magnetic field and temperature at which there
exists a well-correlated liquid of field-induced vortices,
the dc conductivity as a function of increasing wavevector, $k$, was
argued to increase for small values of $k$ and then decrease at large
values of $k$.  The increase in the conductivity at small $k$ arises
from viscous drag between vortices\cite{m&n} which impedes their 
relative motion and therefore decreases their contribution to 
resistance in a nonuniform current.  However, when the length scale
of the nonuniformity in the current is smaller than the intervortex
spacing (high $k$), the conductivity is more determined by the
short-distance correlations of the superconducting order parameter,
rather than the vortex interactions.  In this short-distance regime
the behavior is as in zero magnetic field: the conductivity
decreases with increasing $k$.

Because this effect is only expected to appear in a regime with 
strong correlations and fluctuations, analytical calculations 
do not appear feasible.  In the present paper we report on
computer simulations of the time-dependent Ginzburg-Landau (TDGL)
equation in two dimensions which allow us to study the nonlocal dc
conductivity both in real space and in wavevector space as a 
function of magnetic field, $H$, and temperature, $T$.  We have 
indeed observed 
the nonmonotonic $k$-dependence in the vortex-liquid regime, 
as expected.\cite{us,bmp}  

To measure the nonlocal transport properties experimentally 
requires applying currents that are nonuniform on the appropriate 
length scale.  This has been done in two types of experiments, 
so far.\cite{tvk}  Israeloff, {\it et al.}\cite{isr} have made 
$3 \mu$m-size
wire loops, applied the current asymmetrically, and measured
resistances as a function of a magnetic flux passing through
the loop.  Safar, {\it et al.}\cite{safar} have applied contacts
to both sides of $10-50 \mu$m-thick YBCO samples, measuring
effects due to the nonuniformity of the current across the
sample.  This latter experiment detected the nonlocal resistivity
along the direction parallel to the vortex lines, due to the
lines having integrity (not breaking) across the sample in
a portion of the vortex-liquid regime.  The present work indicates 
that the phenomenon is quite general, and could be studied in
superconductors of any dimensionality.  What is required is to
be able to apply nonuniform currents and measure voltages on 
length scales of order the appropriate correlation length of 
the superconductor.  With modern microfabrication techniques 
(and/or possibly using scanning-tip probes) this should
be feasible for a broader range of materials
and geometries than those used in the two experiments discussed
above.  What would be best would be an experiment that
simultaneously and quantitatively probes a range of length
scales, so the dependence of the transport properties on
length scale (and other parameters) can be systematically
studied.\cite{s&k}  We hope the results reported below 
help motivate such studies.  

An outline of this paper is as follows.
Section II describes the simulation.  Section III outlines the phase
diagram of a two dimensional superconductor, providing a context for
our results.  Section IV describes the conductivity we observe in 
both real and wavevector space in the different regimes of the phase
diagram.  Finally, in Section V, the results are discussed, both 
in the context of earlier work and in terms of possible experiments.

\section{The Simulation}

We wish to study the nonlocal conductivity of a two-dimensional 
(thin-film) sample of a strongly type-II superconductor.  
We begin with the
time-dependent Ginzburg-Landau (TDGL) equation (MKSA units):
\begin{equation}
\Gamma^{-1} (\partial_{t} + i {e^{*}\over \hbar} \Phi) \Psi =
{\hbar^{2} \over 2 m^*} ({\bf \nabla} 
- i {e^{*}\over \hbar} {\bf A} )^{2} 
\Psi - a \Psi - b|\Psi|^{2} \Psi + \zeta({\bf r},t).
\label{tdgl1}
\end{equation}
$\Psi({\bf r},t) = \psi({\bf r},t) e^{i\phi({\bf r},t)}$ 
is the superconducting order parameter.
$\Gamma$ is the kinetic coefficient for the relaxation of the order 
parameter towards equilibrium, it is assumed to be real.
$m^*$ is the effective mass of a Cooper pair and 
$e^{*}=2e$ is the charge of a Cooper pair.
The noise, $\zeta$, is Gaussian distributed and
$\delta$-function correlated, with the coefficient given by the
fluctuation-dissipation theorem:
\begin{equation}
\langle \zeta^{*} ({\bf r},t)\zeta ({\bf r'},t')\rangle 
= 2 \Gamma^{-1}
k_{B} T \delta^{(d)}({\bf r-r'}) \delta(t-t').
\label{noise1}
\end{equation}
$\Phi$ and ${\bf A}$ are the scalar and vector potentials,
respectively.  We consider the strongly type-II (large $\kappa$) 
limit, where the order parameter fluctuations are much stronger
than the magnetic field fluctuations.  Thus we use a uniform, 
static magnetic field.    

To simplify this equation we rescale:  Energy is measured in units 
of $|a|$. We work only below the mean field $T_c$, so we set $a=-1$.  
The order parameter magnitude is measured in units of 
$\sqrt{|a| / b}$, 
its equilibrium value at zero noise and zero magnetic field;  
therefore, $b=+1$.  
Length is measured in units of $\xi={\hbar \over \sqrt{2m^*|a|}}$, 
the order-parameter correlation length at zero noise and zero 
magnetic field, so ${\hbar^2 \over 2 m^*}=1$.
Magnetic flux is measured in units such that the flux quantum 
is $2 \pi$;  that is ${e^* \over \hbar} =1$.  
Time is measured in units of ${1 \over \Gamma |a|}$, 
so $\Gamma =1$.  We also take $k_B=1$.  The parameters remaining 
are the temperature, now measured in units of $|a|$; 
and the magnetic field, in units of flux quanta per area $2\pi\xi^2$, 
or, equivalently, in units of the mean-field upper critical field, 
$H_{c2}^{MF}(T)$.  
The theory also needs an ultraviolet cutoff, which we realize in the
simulations by discretizing space.  Note that when our rescaled
temperature is large ($T>>1$), this means $k_BT >> -a >0$, which is 
the regime of strong thermal fluctuations {\it below} the 
mean-field transition temperature, $T_c^{MF}$.

The rescaled equation is
\begin{equation}
(\partial_{t} + i \Phi) \Psi =
(\nabla - i {\bf A} )^{2} \Psi
+ \Psi - |\Psi|^{2} \Psi + \zeta.
\label{tdgl2}
\end{equation}
To solve this equation, we discretize space and time.  The film is 
approximated by a square lattice with spacing one in the rescaled 
units.  Time is divided into a series of time steps.  
The length of these time steps must be decreased at higher 
temperatures in order to obtain accurate steady-state results and
avoid numerical instabilities.  We used time steps in the range of 
0.2 to 0.02 rescaled time units for temperatures ranging from 0 to 
1, respectively.
   
Taking advantage of the gauge invariance of the equation, we work in 
terms of only gauge-invariant quantities, namely, the order parameter 
magnitudes at each site of the lattice, $\psi({\bf r})$, and 
the gauge-invariant phase differences along each nearest-neighbor
link, $\theta$, defined by
\begin{equation}
\theta({\bf r,r'}) = \phi({\bf r'}) - \phi({\bf r}) 
- \int_{\bf r}^{\bf r'} {\bf A} \cdot d{\bf \ell} ,
\end{equation}
where ${\bf r}$ and ${\bf r'}$ are adjacent lattice points and the 
integral is along the straight line between them.
Derivatives are approximated using differences, 
and only differences up to one time step and two
lattice spacings are kept; this is the minimum needed to
approximate the derivatives appearing in the TDGL equation.  

We use periodic boundary conditions in both the $x$ and $y$ 
directions, giving effectively a toroidal surface.  We begin with
random initial conditions, setting the gauge-invariant phase
difference to a random number between $-\pi$ and $\pi$ on three of 
the four links surrounding each plaquette; the phase difference on 
the last link is then determined uniquely (modulo $2\pi$) by the 
magnetic flux passing through the plaquette.  
The order parameter magnitude begins at the small, spatially 
uniform value of 0.01.

Our goal is to study the conductivity.  To do this we apply an
electric field and measure the electrical current.  The TDGL equation
in this type-II limit only involves the supercurrent due to the
paired electrons that produces enhanced conductivity over that
of the normal state.  The supercurrent, in our rescaled units, is
\begin{equation}
{\bf J_s} =  2 Im \{ \Psi^* ({\bf \nabla}
- i {\bf A}) \Psi \}.
\end{equation}
This we also discretize as described above, so the supercurrent is
defined on each nearest-neighbor link of the lattice as
\begin{equation}
J_s({\bf r,r'}) = 2 \psi({\bf r})\psi({\bf r'})
sin\bigl(\theta({\bf r,r'})\bigr) .
\end{equation}
In addition to this supercurrent, there is also a normal current
which we will assume is simply local and Ohmic: 
${\bf J_n(r)} = \sigma_n{\bf E(r)}$, where $\sigma_n$ is the
normal-state conductivity.

We generally apply an electric field that is uniform in the $y$ 
direction, parallel to ${\bf \hat x}$ and given by a 
$\delta$-function along $x$:
\begin{equation}
{\bf E} = E_o \delta (x) {\bf \hat x}  .
\end{equation}
On the lattice this $\delta$-function is realized by applying the
electric field only between two adjacent columns of lattice sites.
In this geometry the resulting current is also parallel to 
${\bf \hat x}$ and, in the linear response regime is given by
\begin{equation}
J(x) = E_o \int dy \sigma_{xx}(x,y) = E_o \sigma_{xx}(x,k_y=0).
\end{equation}
Thus we measure the dependence of the nonlocal conductivity on the 
spacing $x$ for conditions that are uniform along $y$ ($k_y = 0$).  
By fourier transforming on $x$ we also obtain 
$\sigma_{xx}(k_x,k_y = 0)$.

\section{Phase Diagram}

Our simulation is designed to represent a superconducting film with 
thickness less than the bulk order-parameter correlation length.  
When there is 
no applied magnetic field, such a system is expected to undergo a
Kosterlitz-Thouless transition at a temperature $T_{KT}$.\cite{KT}
Below this temperature there are bound 
pairs of vortices, but no free vortices and therefore no Ohmic 
resistance to a uniform dc current.  Above $T_{KT}$, there are 
free vortices, whose mobility results in a nonzero resistivity.  

When a magnetic field is applied perpendicular to the plane of such 
a film, the flux is not expelled.  At low temperatures, a triangular 
vortex lattice forms.  Above the melting temperature, $T_M(H)$, this 
lattice becomes unstable to dislocations and melts.  
At intermediate magnetic fields -- i.e. much less than the mean-field
$H_{c2}^{MF}(T)$ but greater than one flux quantum per magnetic
penetration length squared -- $T_M$ is expected to be only weakly
field-dependent.\cite{fisher}  In the limit of a normal-state sheet
resistance much larger than the quantum of resistance, 
${\hbar \over (e^*)^2}$, 
$T_M = (0.026 \pm 0.008) T_{KT}$ in this intermediate field 
regime.\cite{fisher}
When temperature is rescaled as we have done, this relation between
$T_M$ and $T_{KT}$ holds for films with smaller sheet resistance as 
well.  At higher magnetic fields, near mean-field $H_{c2}^{MF}(T)$,
$T_M \propto (H_{c2}(0)-H)^2$ for Ginzburg-Landau theory 
with thermal fluctuations.\cite{fisher}  The locations of the
phase transitions are roughly sketched in Figure 1.

To estimate the zero-field transition temperature, we used 
a finite-size scaling analysis of the order
parameter phase correlations, obtaining $T_{KT} = 0.8 \pm 0.1$,
which implies $T_M = 0.02 \pm 0.01$ in the intermediate field regime.
We did not attempt to directly estimate $T_M$ by looking for the
melting transition in our simulations.
We also raised the applied field until the zero-temperature 
equilibrium order-parameter magnitude dropped to zero to find that 
$H_{c2}(T=0)=1.18 \pm 0.02$.  In the continuum, $H_{c2}(T=0)=1$ in 
our rescaled units; the increase to roughly $1.2$ is due to our 
approximating the continuum by a lattice with spacing $\xi$ and 
approximating the spatial derivative with the lowest-order 
difference.  We expect other quantities are also quantitatively 
shifted small amounts by these approximations.  
In particular, the precise values
we obtain for the conductivity are likely to be affected.  
However, as we are focusing on trends related to the variation of 
$x$, $k$, $H$ and $T$ rather than on precise numerical values, 
these shifts are not expected to affect our conclusions.

\section{Results}

We studied points in the phase diagram in 
the field range $0 \leq H \leq 1.2$ 
and the temperature range $0 \leq T \leq 1$ as shown in Figure 1.    
At most points we studied, all the vortices present were
field-induced.  Thermally-induced vortex-antivortex pairs 
were observed only for $T \geq 0.1$.  However, the number 
of field-induced vortices was much
greater than that of thermally-induced vortices for temperatures
up to 0.5 at the nonzero field values studied. 

We saw three characteristic behaviors for the nonlocal
conductivity.  The points A, B, and C shown in Figure 1
are chosen as clear examples of these behaviors.  Point A, at T=0.001
and H=$2 \pi /25$, is representative of what is seen in the ordered
phases at very low
temperatures and fields, including T=0 and H=0.  Point B, at T=0.04
and H=$ \pi /5$, is representative of the well-correlated 
vortex-liquid regime, intermediate in both
temperature and field.  
vortices.  
Finally, point C, at T=0.1 and H=$6 \pi /25$, is representative of
high temperature and high field behavior.  

In linear response, the nonlocal conductivity in real space,
$\sigma_{xx}(x,k_y=0)$, for the characteristic points are 
as follows (Figure 2):  At point C, we see a
conductivity which is sharply peaked around $x=0$, falling off
exponentially to zero with a length scale of order the correlation
length.  This behavior is in qualitative agreement with the
high-temperature behavior obtained analytically from lowest-order
fluctuations about the mean-field normal state.\cite{us}  
At point B the conductivity
still has a sharp peak at $x=0$ of similar width to that at point C,
but it then drops {\bf below} zero over a distance of few
intervortex spacings before returning to zero.  This is the
negative nonlocal conductivity expected in this well-correlated
vortex-liquid regime.\cite{us}  At both points B and C the
conductivities and correlation lengths are all finite, so the
numerical results for a finite sample that is large compared
to all correlation lengths do not show finite-size effects and
thus are a faithful representation of a much larger sample.  

In the ordered phases the uniform conductivity is infinite.  In the
Meissner phase ($H=0$) this is true for the continuum system as well,
while for the vortex lattice phase the infinite uniform ($k=0$) 
conductivity is due to weak pinning of the vortices to our
numerically-imposed lattice that impedes the ``flux-flow''
that would occur for the continuum system.  To avoid this divergence,
instead of applying only the $\delta$-function electric field (2.7) 
as was done at points B and C, at point A we also apply a 
compensating spatially uniform electric field that cancels the 
uniform ($k=0$) component 
coming from the $\delta$-function.  This means we do not measure the
very large $k=0$ part of the conductivity and the data for point A
in the inset to Figure 2 consequently has an 
arbitrary $x$-independent vertical shift.
The conductivity vs. $x$ shows a broad peak at the origin, with a 
width and magnitude proportional to the linear sample size 
and a shape well fit by a parabola centered on the column 
opposite to where the $\delta$-function component of the
electric field is applied.

Still in linear response, but now in $k$-space (see Figure 3), at
point C, the conductivity $\sigma_{xx}(k_x,k_y=0)$ falls 
monotonically from its value at $k_x=0$.  At point B
the conductivity rises from its value at $k=0$ to a maximum at a
wavevector roughly corresponding to the inverse of the vortex 
spacing and then declines.  
At point A (inset) the conductivity falls off
monotonically like $1/k^2$, as expected.\cite{us} 

At point B, when the system
is driven beyond linear response, the effect is to reduce the
magnitude both of the peak in $\sigma(x)$ and of the negative 
regions.  In $k$ space, the effect is to suppress the 
nonmonotonicity.  Figure 4
shows $\sigma (k)$ for a series of applied electric fields of
increasing magnitude.  Here in the nonlinear regime we apply the
$\delta$-function electric field (2.7), measure $J_s(x)$, define the
nonlocal and nonlinear conductivity as $\sigma(x)=J_s(x)/E_o$, and
show its fourier transform in Figure 4.  The higher electric fields
produce a strong shear flow in the vortex liquid, which apparently
reduces the effective viscosity of the vortex liquid.  

For points in the vortex liquid below the dashed line in Figure 1,
the nonlocal conductivity in $k$-space showed nonmonotonicity 
significantly outside the statistical errors.
Above this line nonmonotonic behavior is still observed at many
points; however, it is not possible to distinguish between true
nonmonotonicity and statistical noise at these points, 
given the level of accuracy of the present data.
Thus the dashed line is actually a lower bound on the
true boundary between monotonic and nonmonotonic $\sigma(k)$.
This boundary is only a crossover, not a phase transition; 
it is roughly where the correlation length of the conductivity 
(the range of the nonlocal conductivity in real space) 
becomes comparable to the intervortex spacing.
An interesting question is whether this boundary, like
the melting line, intersects the zero-field axis at a temperature
below $T_{KT}$.  Following this line to lower fields in our 
simulations is difficult.  
It goes to higher temperatures where smaller
time steps are required for numerical stability.  Also, to study
lower fields, larger samples are required in order to include a
sufficient number of field-induced vortices  
and to probe distances well beyond the intervortex spacing.  
Following the line to lower temperatures is also difficult.  As it
approaches the melting line, nonlinear response sets in at
progressively lower applied electric fields, and longer runs are
required in order to obtain good signal to noise ratios in the 
linear response regime. 

\section{Discussion}

To discuss the phenomena we've observed, and in particular to 
connect with experimental possibilities, 
it is useful to speak in terms of
resistivity.  $\rho ({\bf k})$ is simply $\sigma({\bf k})^{-1}$, 
and $\rho({\bf r})$ is its fourier transform.  
\begin{equation}
E_{\mu}({\bf r})=\int \rho_{\mu \nu}({\bf r}-{\bf r'})
J_{\nu}({\bf r'}) d{\bf r'}  .
\end{equation}
When $\sigma_n << \sigma_s$, $\rho(k)={1 \over \sigma_n 
+ \sigma_s(k)}
\sim \rho_s(k) - \sigma_n [\rho_s(k)]^2$, and when $\sigma_n >>
\sigma_s$, $\rho(k) \sim \rho_n - \sigma_s(k)\rho_n^2$, where 
$\rho_s(k)={1 \over \sigma_s(k)}$ and $\rho_n={1 \over \sigma_n}$.

The resistivity in real space, $\rho_{xx}(x,k_y=0)$, for the case of 
very low normal-state conductivity is sketched in Figure 5 using the 
nonmonotonic $\sigma_s(k)$ at point B from Figure 3.  This shows the
response of a film in the $xy$ plane in a magnetic field parallel to
$z$ to a current applied between two closely-spaced line contacts 
parallel to each other and to the $y$-axis, in the geometry shown 
in Figure 6.  If the current contacts are sufficiently closely spaced 
to approximate well a $\delta$-function in current, the resulting
electric field pattern is proportional to $\rho_{xx}(x, k_y=0)$.  
The effect is as follows:
When a current is applied in the $x$ direction between the two line
contacts, the vortices which are positioned between the contacts feel
a force in the $-y$ direction.  When they move in response to this
force, the vortices near to but outside the region in which current 
is flowing are dragged along.  An electric field will be seen between 
the two current contacts due to the phase-slip caused by the moving 
vortices as well as simply the flow of normal current.  
In addition, an electric
field of the same sign will be present {\it outside} the region 
in which current is flowing, due
to phase-slip generated by the vortices being dragged by their
neighbors.  The negative regions in the nonlocal resistivity at short
distance are due to the supercurrent being nearly uniform on length
scales shorter than the order-parameter correlation length.  
Thus to get zero net current just
outside of the current contacts in Figure 6 a counterflowing normal
current must be present to cancel the supercurrent there.
This results in a negative nonlocal resistivity for distance $x$
shorter than the intervortex spacing, as seen in Figure 5.

Returning to the conductivity measured in our simulation, the 
negative regions observed in Figure 2 for point B represent places 
where the current must flow in the opposite direction to that 
of the applied electric field in order to keep vortices 
stationary which would otherwise be dragged along by their
moving neighbors.  These negative regions in $\sigma(x)$, at $x$ of 
order the intervortex spacing or more, correspond to the positive 
regions in $\rho(x)$ for the same $x$ range, and both represent 
the effect of viscous drag between vortices in the liquid state.

To connect to the work in our earlier paper,\cite{us} in which
$\sigma({\bf k})$ was expanded as 
\begin{equation}
\sigma_{\mu \nu}({\bf k})=\sigma_{\mu \nu}(0)
+S_{\mu \alpha \beta \nu} k_{\alpha} k_{\beta},
\end{equation}
our results indicate that $S_{xxxx}$ is not always 
negative.\cite{bmp}  At
intermediate values of the temperature and field, $S_{xxxx}$ is {\it
positive}, corresponding to a conductivity  that increases with
increasing wavevector $k$ at small values of $k$.  

Furthermore, we showed previously that, when a hydrodynamic treatment 
is appropriate, the elements of the $S$ 
tensor can be written as linear combinations of the
elements of the vortex-liquid viscosity tensor defined in the 
theory of Marchetti and Nelson.\cite{m&n}  Specifically, the element 
$S_{xxxx}$ is
proportional to the viscosity due to the variation in the $x$
direction of vortex motion in the $y$ direction.  
A positive $S_{xxxx}$
corresponds to a positive vortex-liquid viscosity in a 
well-correlated liquid of field-induced vortices. 

However, the negative $S_{xxxx}$ values obtained in the
high $T$ and/or high $H$ regime should not be thought of as
negative vortex-liquid viscosities.\cite{b&m}  
$S$ does not represent a vortex-liquid viscosity when the system 
cannot be described hydrodynamically.    
At high temperatures and/or fields, the correlation length of the
conductivity is
shorter than the spacing between vortices, and the value
of $S$ comes from interactions not on the scale of vortices but 
on the shorter scale of the order-parameter correlation length.  
The conductivity decreases with increasing wavevector as in
zero magnetic field.  One view of this regime is that the 
vortex-liquid viscosity becomes so small that the other, 
shorter length scale contributions to the nonlocal conductivity 
dominate, even at small $k$.

At intermediate fields and/or temperatures when a hydrodynamic 
picture is applicable, there are actually two regimes:
Here we have presented data at intermediate temperatures and 
nonzero field where there is a well-correlated liquid of 
field-induced vortices.  
In this case the positive vortex-liquid viscosity causes moving 
vortices to drag along other parallel vortices, thereby producing 
a positive nonlocal resistivity for $x$ of order the intervortex 
spacing or larger, as in Figure 5.  
At temperatures just above the Kosterlitz-Thouless
transition in zero field, on the other hand, there is a neutral 
liquid of thermally-induced vortices {\it and antivortices}.  
In this regime, a moving vortex is more likely to drag a nearby
{\it antivortex}, which is attracted to it, than it is to drag one
parallel to itself, so the usual positive vortex-liquid
viscosity instead produces a negative nonlocal resistivity.  
We have not used our simulation to study this latter regime.

It appears that experimental observation of these nonlocal transport
properties of type-II superconducting films is feasible.  
What is needed?
A high normal-state sheet resistance, $R_n$, 
would maximize the extent 
of the fluctuation regime above the melting line, 
and it would also minimize the
contribution of normal currents to measured quantities.  
Vortex pinning should be weak enough so that the vortex-liquid 
regime of interest is not strongly pinned.  
Many different sample geometries could satisfy the basic 
requirement that the applied current vary on a length scale of 
order the correlation length of the conductivity.  
Point contacts in the middle or at the edge of the film would
produce two-dimensional current patterns.  Line contacts allow
one-dimensional patterns, including approximations to step-function  
or delta-function configurations.  
The latter is closest to what we have studied in our simulation.
Figure 6 shows an arrangement of a series of 
line contacts parallel to one another with spacings of order the
intervortex spacing (or, for $H=0$, of order the order-parameter
correlation length).
Current would be run between nearest-neighbor contacts 
J1 and J2, and the voltage differences would 
be measured between the remaining contacts where there is 
no net current flowing.  Alternatively, a step-function applied
current pattern could be made by passing current between two more 
widely separated line contacts in Figure 6.

The three regimes would appear as follows:  (C) In the 
high-temperature
or high-field regime, including the normal state, 
the voltage difference between all V contacts not between
the current contacts will
be zero, because the resistivity is local on the length scales
studied and no current flows there.  However, if the contact spacing
is less than or of order the order-parameter correlation length, 
and well below
the intervortex spacing, then the negative nonlocal resistivity
should be seen as a negative electric field that falls off
as one moves away from the current contacts.  This latter regime
should be what is seen on approaching $T_c$ from above for
zero magnetic field.  (A) In the low-temperature, 
low-field regime, below the melting line, assuming the pinning is 
negligible, the linear-response electric field measured between each 
pair of neighboring V
contacts will be equal and nonzero due to the uniform motion of the
vortex lattice induced by the current.   This is also assuming the
current contacts are centered on the sample so the net torque on the
vortex lattice vanishes.  (B) In the well-correlated vortex-liquid
regime, the electric field a vortex spacing or more from
the current flow will be positive, as the vortices in this region 
are dragged by their neighbors.
This electric field will decrease with a vortex-liquid correlation 
length as one moves away from where the current is flowing.

To estimate the magnitude of the voltages expected, we need a value 
for the factor ${e^* \over \hbar \Gamma |a|}$ by which we have 
rescaled the electric potential.  
An estimate of $\Gamma |a|$ can be obtained by
comparing the theoretical expression for the fluctuation 
conductivity calculated using the time-dependent Ginzburg-Landau 
equation,
\begin{equation}
\sigma=k_B T {e^2 \over \hbar^2} {1 \over 2 \pi \Gamma |a|}
\end{equation}
(in two dimensions), with experimentally obtained values, 
as shown in the review by Skocpol and Tinkham.\cite{s&t}
The result is a voltage scale of order $10^{-4} V$ for materials 
with bulk transition temperatures of order one Kelvin.  
This value is the same as that obtained by dividing the 
characteristic energy $k_B T_c$ by the electron charge.  
To obtain linear response at point B we had to decrease the voltage
one to two orders of magnitude below this scale, 
thus to the order of microvolts.

As for the current, the factor by which it has been rescaled is
${|a|^{3/2} \over b}{e^* \over \sqrt{2 m^*}}$.  A rough
estimate of its scale can be obtained by dividing the voltage scale
by the quantum of resistance and by the order-parameter 
correlation length.  
The result is in the range $(10^{-2}-1) A/cm$.  Thus the relevant
current and voltage scales appear to be well within the reach of
present experimental technique (although combining this with the
small length-scales required may be quite a challenge).

To conclude, our simulation has displayed negative nonlocal
conductivity in a vortex-liquid regime, which is distinct from the
behavior of the vortex lattice and the higher temperature region.  
We believe that this signature of the well-correlated 
vortex liquid is experimentally
observable and could provide useful information on vortex dynamics
and interactions in type-II superconductors. 

\acknowledgements 

R.W. gratefully acknowledges support from AT\&T as well as helpful
discussions with A.J. Leggett and S. Sondhi.  D.H. thanks
L.I. Glazman and M.A. Moore for discussions.

\clearpage

\begin{center}
{FIGURE CAPTIONS}
\end{center}
{\sc FIG. 1}. Sketch of the phase diagram of the model thin-film
superconductor we have simulated.  Note the logarithmic scale on 
the temperature axis.  The solid curve is a rough estimate of the
melting boundary based on our estimate of $T_{KT}$ (marked with a
diamond) and Ref. 12.
The points we have studied are marked with crosses, and
representative points A, B, and C are marked with squares.  
Negative nonlocal conductivity was observed for the points
below the dashed line.  The near intersection of the dashed
line and the (solid) melting curve may not be significant,
because both have large uncertainties in their precise
locations (see text).

{\sc FIG. 2}. Conductivity $\sigma_{xx}(x,k_y=0)$ as a function of 
position $x$ at the points B and C indicated in the phase diagram
(Fig. 1).  Note the negative nonlocal conductivity for B.
Inset also includes point A; the vertical zero is shifted 
for the data from point A only (see text).  The local normal-state
contribution to the conductivity at $x=0$ has not been included here.

{\sc FIG. 3}. Conductivity as a function of wavevector at the points
B and C.  Inset includes point A.  These data are simply the fourier
transforms of those in Fig. 2, so the $k$-independent normal-state
contribution is not included.

{\sc FIG. 4}. Nonlinear conductivity (see the text for a precise 
definition) as a function of wavevector in the vortex liquid at 
the point B for five values of electric field:  0.0125, 0.025, 
0.05, 0.1 and 0.2.

{\sc FIG. 5}.  Nonlocal resistivity $\rho_{xx}(x,k_y=0)$ as a 
function of position $x$ at the point B in the limit of low 
normal-state conductivity.

{\sc FIG. 6}.  An idealized proposed experimental geometry.  Shown
is a superconducting film with parallel line contacts spaced by a 
distance less than or of order either the intervortex spacing or 
the correlation length of the superconducting order parameter.  
The applied current 
is run from J1 to J2, and voltage differences between the V contacts 
are measured.

\clearpage

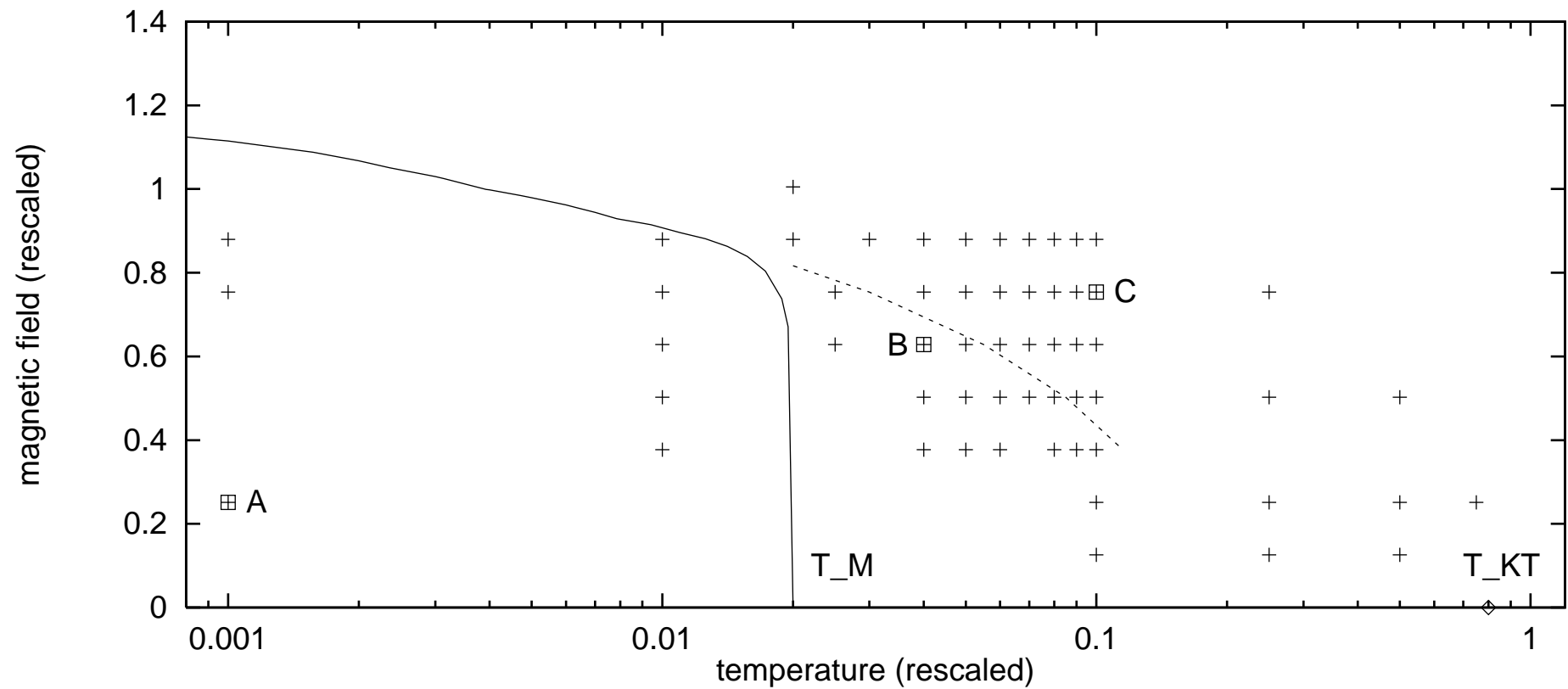
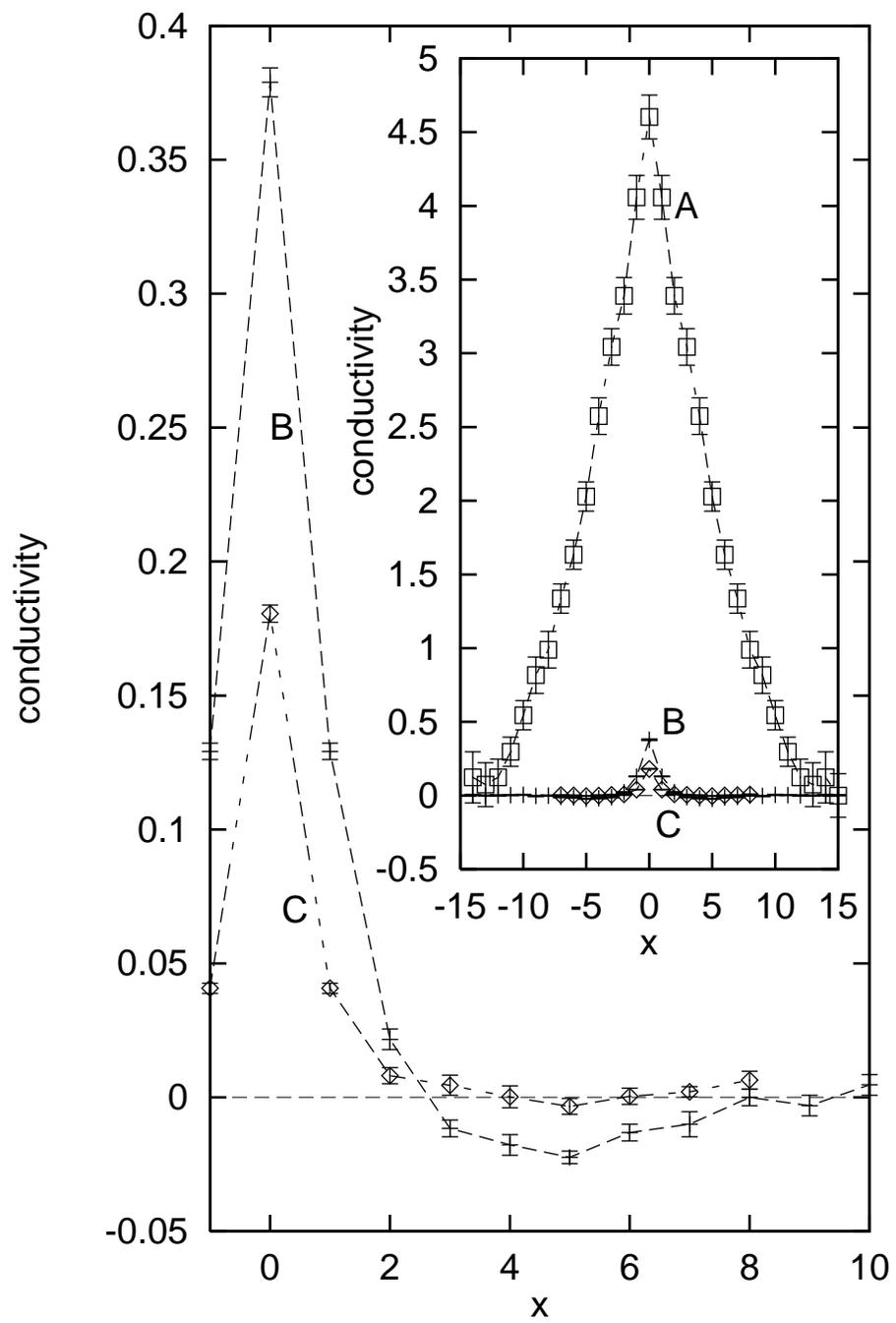
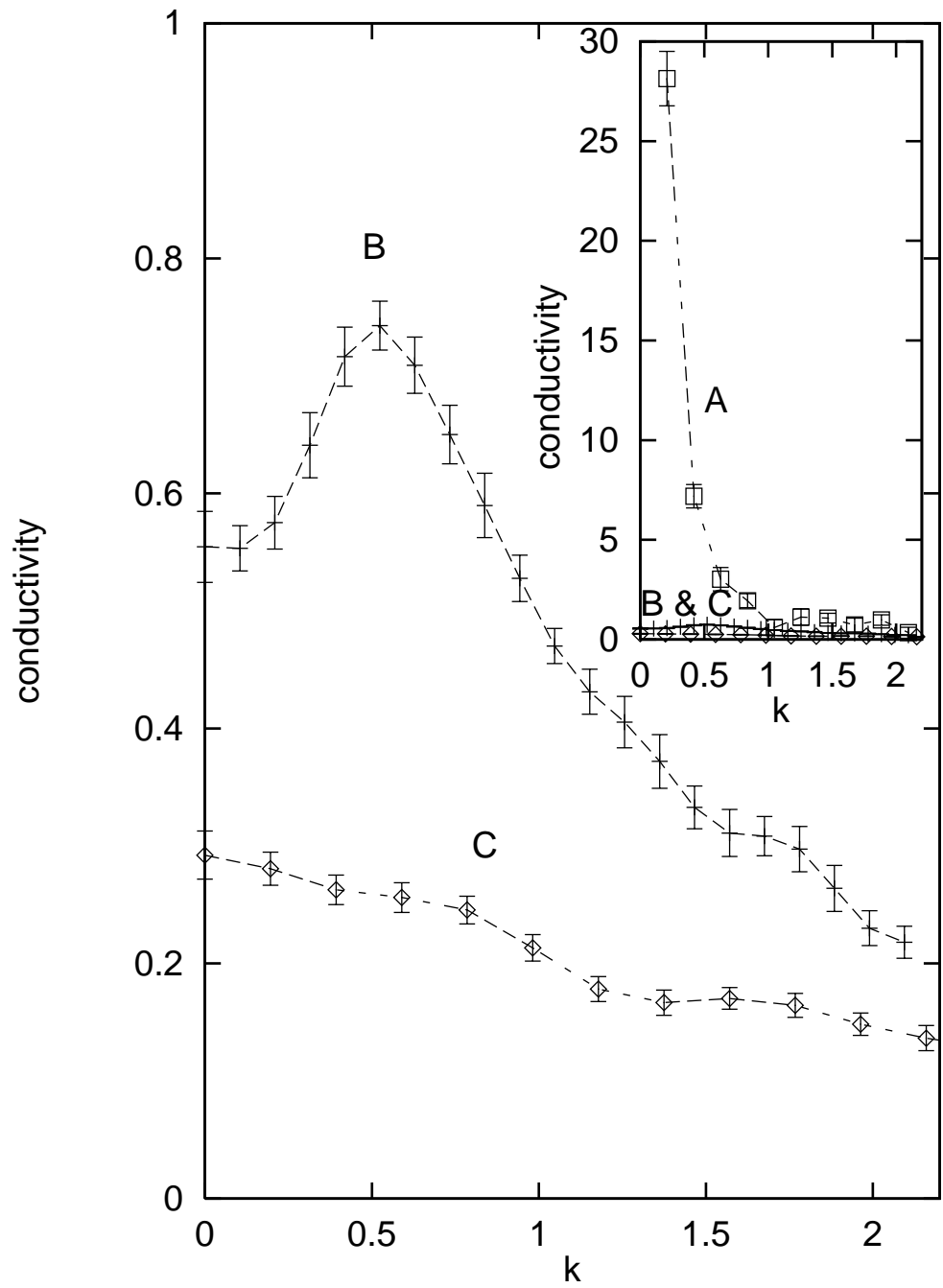
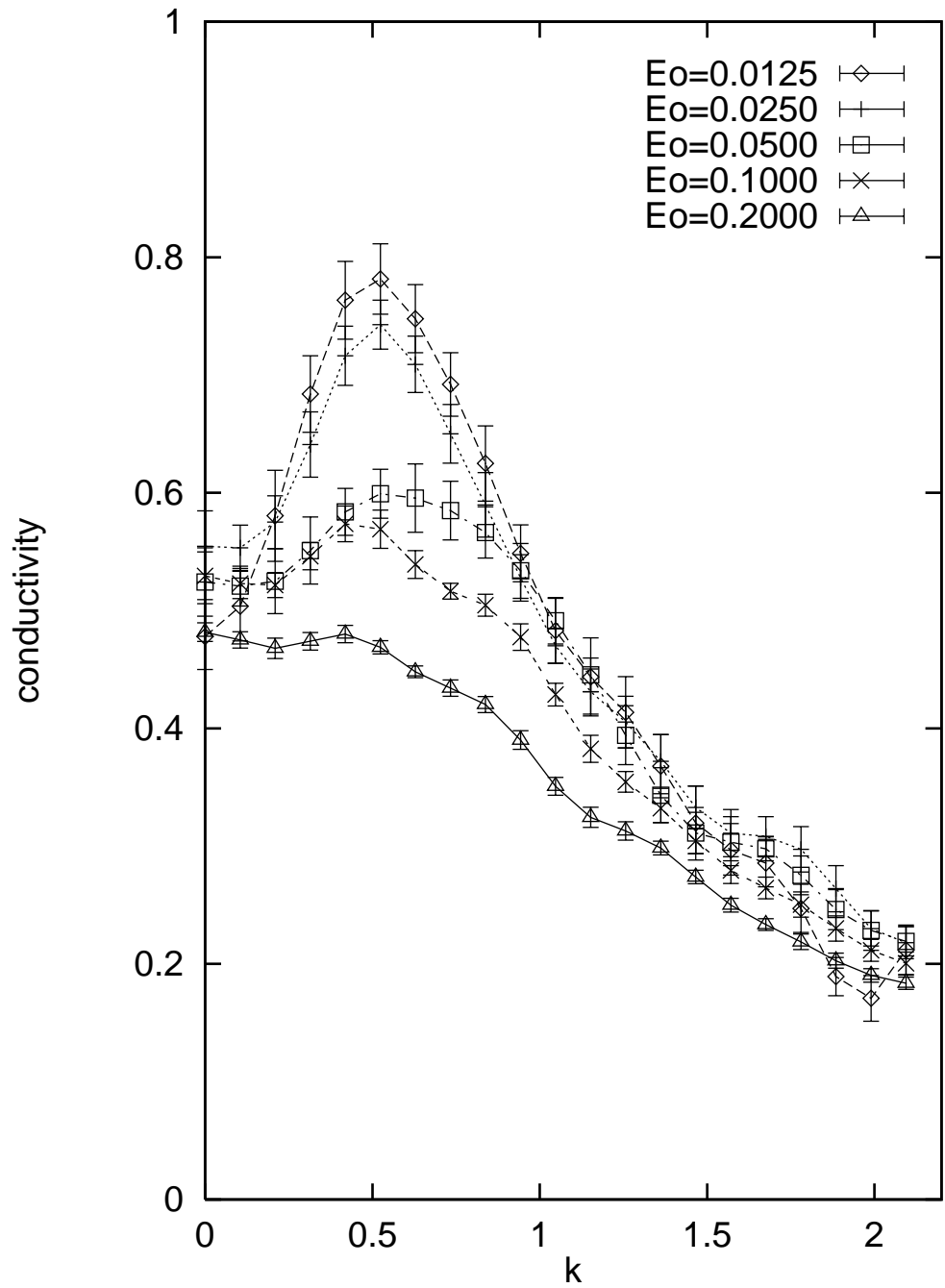
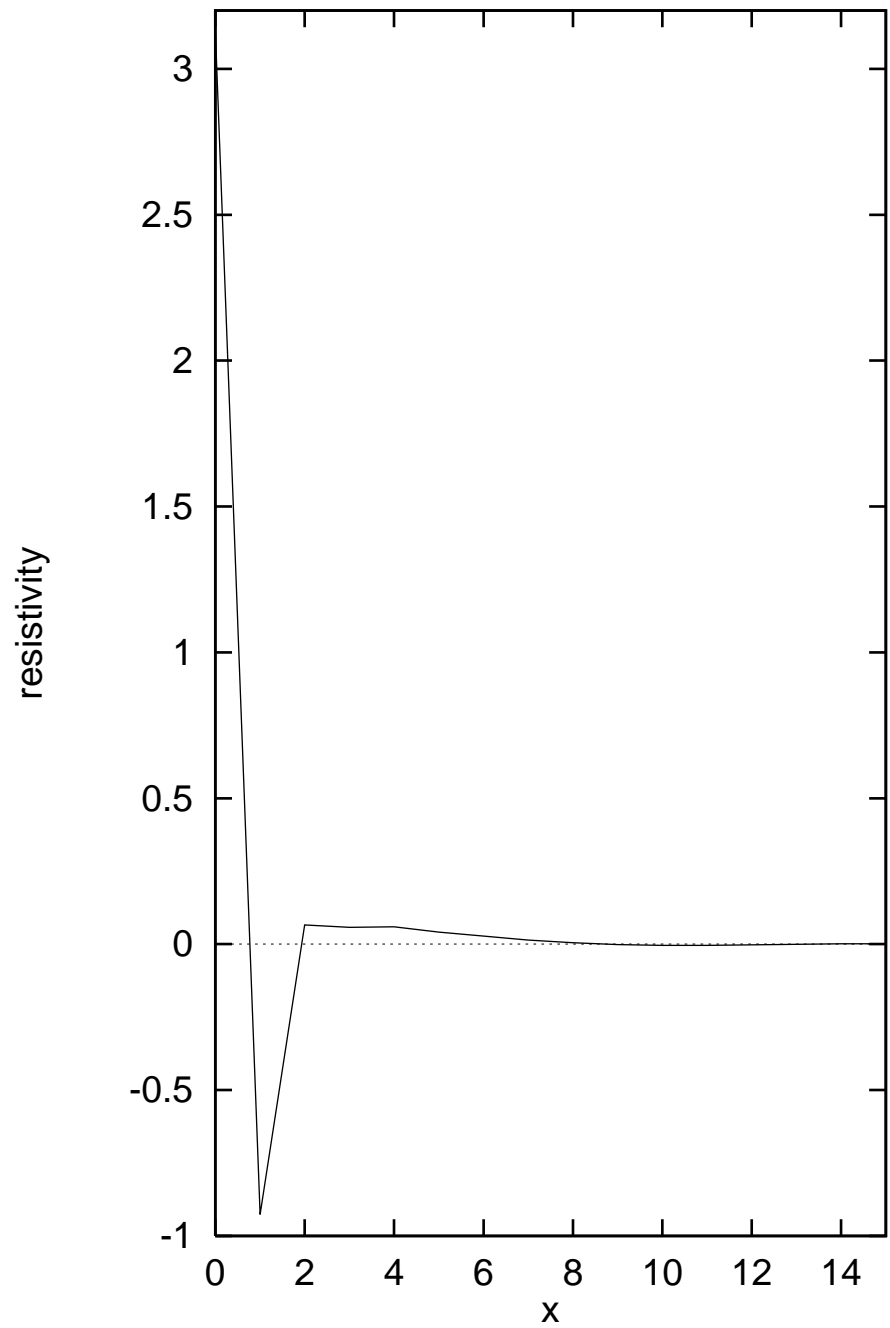
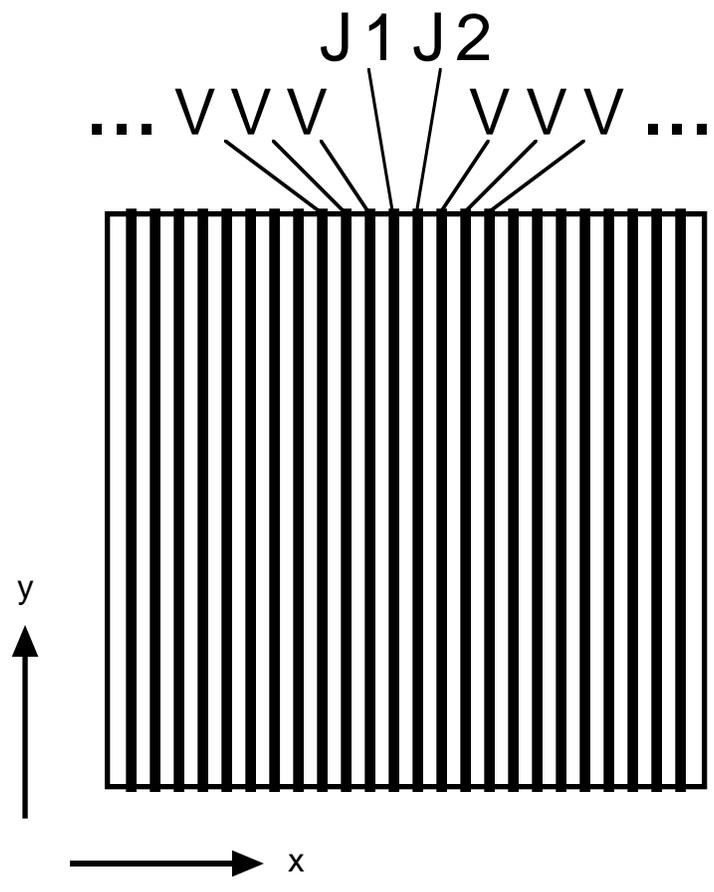
\end{document}